\documentclass[prb,showpacs,amsmath,amssymb,superscriptaddress,twocolumn,floatfix]{revtex4-1}

\usepackage{graphicx}
\usepackage{color}
\usepackage{epstopdf}
\usepackage{mathrsfs}  

\def\unit#1{\ \mathrm{#1}}
\def\zz{\mbox{\v Z}}
\def\diag{\mbox{diag }}

\def\Heff{\hat{H}_{\rm eff}}
\def\eins{1\!\! 1}  
\def\magDir{\mathbf{e}_M}

\begin{document}

\title{Magnetic anisotropies of quantum dots}
\author{Karel V\'yborn\'y}
\affiliation{Department of Physics, University at Buffalo--SUNY,
Buffalo, New York 14260, USA}

\affiliation{
Institute of Physics, ASCR, $v.~v.~i.$, 
Cukrovarnick\'a 10, CZ-16253 Praha 6, Czech Republic}

\author{J.~E. Han}
\affiliation{Department of Physics, University at Buffalo--SUNY,
Buffalo, New York 14260, USA}

\author{Rafa\l{}  Oszwa\l dowski}
\affiliation{Department of Physics, University at Buffalo--SUNY,
Buffalo, New York 14260, USA}

\author{Igor \v Zuti\'c}
\affiliation{Department of Physics, University at Buffalo--SUNY,
Buffalo, New York 14260, USA}

\author{A.~G. Petukhov}
\affiliation{South Dakota School of Mines and Technology, 
Rapid City, South Dakota 57701, USA}

\date{Feb14, 2012}

\begin{abstract}
Magnetic anisotropies in quantum dots (QDs) doped with magnetic ions are
discussed in terms of two frameworks: anisotropic $g$-factors and
magnetocrystalline anisotropy energy. It is shown 
that even a simple model of zinc-blende p-doped QDs displays
a rich diagram of magnetic anisotropies in the QD parameter space. 
Tuning the confinement allows to control magnetic easy axes in QDs 
in ways not available for the better-studied bulk.
\end{abstract}



\maketitle

\section{Introduction}

Once the origin of magnetic ordering in a specific material is
understood, it is often important to determine its magnetic anisotropy (MA)
and hard and easy magnetic axes in particular. A shift of focus towards MA has
already occurred for the studies of bulk dilute magnetic
semiconductors (DMS),\cite{Welp2003:PRL,Liu2003:PRB} 
but not yet fully for magnetic
quantum dots (QDs) where it could play certain role, for example, 
in context of transport phenomena,\cite{Fernandez-Rossier2011:bookCh}
the formation of robust magnetic 
polarons,\cite{Yakovlev2010:bookCh,Beaulac2009:Sci,%
Sellers2010:PRB,Henneberger2010:bookCh} 
control of magnetic ordering,%
\cite{Abolfath2008:PRL,Govorov-twice,Oszwaldowski2011:PRL,Bussian2009:NMat,%
Ochsenbein2009:nnano} nonvolatile memory,\cite{Enaya2008:JAP}
and quantum bits.\cite{LeGall2010:PRB}

In epilayers of (Ga,Mn)As, a prototypical DMS, the magnetocrystalline
anisotropy energy (MAE) has been found to be a significant and often
dominant source of MA\cite{Zemen2009:PRB,Dreher2010:PRB,Linnik2011:arxiv} 
caused by a strong spin-orbit (SO) coupling. It turns out that the
easy axis direction depends on hole concentration, magnetic doping
level as well as on other parameters. For example, when (Ga,Mn)As was
used as a spin injector, the effects of strain (by altering the choice
of a substrate) were responsible for changing the in-plane to
out-plane easy axis.\cite{Zutic2004:RMP} While the 
strong SO coupling\cite{Fabian2007:APS} is also present in p-type
QD of zinc-blende materials doped with Mn, its effect on magnetic
anisotropies will be significantly modified by the confinement.  The energy
levels in such
`nanomagnets,'\cite{Bader2006:RMP,Fernandez-Rossier2007:PRL,Cheng2009:PRB,%
Kyrychenko2004:PRB}
where the Mn-Mn interaction is mediated by carriers,
depend on the magnetization direction $\magDir=(n_x,n_y,n_z)$.  It
is often assumed that the interaction of magnetic moments with holes
in quantum wells (QWs) or, equivalently in flat QDs, is effectively
Ising-like.\cite{LeGall2010:PRB,Lee2000:PRB} Here we quantify this assumption 
and explore MA using two frameworks: (i) an effective two-level Hamiltonian
with a carrier $g$--tensor,\cite{Merkulov1995:PRB} which is widely 
employed also in theory of electron spin resonance, and (ii) MAE,
which is commonly used to study bulk magnets.

While previous studies focused on specific nonmagnetic
QDs\cite{Pryor2006:PRL} and properties sensitive to system details
(such as precise position of magnetic
ions\cite{Cheng2009:PRB,Bhattacharjee2007:PRB}),
we explore more generic magnetic QD models, which can also serve as a
starting point for more elaborate work. We consider a Hamiltonian comprising 
non-magnetic and magnetic parts, 
\begin{equation}\label{eq:01}
\hat{H}=\hat{H}_{QD}+\hat{H}_{ex}.
\end{equation}
The former encodes both QD confinement and SO interaction, which is
prerequisite for magnetic anisotropies, the latter expresses the
kinetic-exchange coupling between holes and localized magnetic
moments. For transparency, we disregard the magnetostatic shape 
anisotropy\cite{shapeAniso2} and  assume that the QD contains a fixed
number of carriers. We mostly focus on the case of a single hole; 
realistically, such system can be a II-VI colloidal\cite{Beaulac2009:Sci}
or epitaxial\cite{Sellers2010:PRB} QD with a photoinduced carrier.
Magnetic moments of the Mn atoms are taken
to be perfectly ordered (collinear) and are treated at a mean-field level.
The magnetic easy axis is then the direction $\magDir$ for which the
zero-temperature
free energy $F(\magDir)$ is minimized. In this article, we take
two different points of view on $F(\magDir)$. On one hand,
we discuss the lowest terms of $F(\magDir)$ expanded in powers of 
the direction cosines of magnetization ($n_x^2+n_y^2+n_z^2=1$),
inspired by the standard `bulk MAE phenomenology' and 
pay special attention to the case of perfectly cubic QDs,
$F(\magDir)=F_0(\magDir)$. The anisotropies in $F_0$ stem purely
from the crystalline zinc-blende lattice.
On the other hand, $F(\magDir)$ acquires
additional terms in systems with less symmetric confinement. 
We therefore discuss
the anisotropic $g$-factors as a useful framework to handle such systems,
e.g. cuboid QDs (orthogonal parallelepiped; extremal cases are a cube
and an infinitely thin slab, i.e., a QW) and show how the expansion
\begin{equation}\label{eq:02}
  F(\magDir)=F_0(\magDir)+AF_1(\magDir)+A^2F_2(\magDir)+\ldots
\end{equation}
can be constructed using powers of $A$ which reflects the anisotropy
in $g$-factors. We begin
by discussing this latter topic in Section~II (quantity $A$ is
defined by Eq.~(\ref{eq:07}) at the end of Sec.~IIA), then proceed to the
phenomenologic (symmetry-based) expansions of $F_0$ in Section~III and
conclude that Section with calculations of $F_1$ in situations that
are beyond the applicability of the $g$-factor framework.

\section{Effective two-level Hamiltonian}

Since $\hat{H}_{QD}$ is invariant upon time reversal, its
spectrum consists of Kramers doublets.\cite{KramersDbl} 
To study the ground-state
energy in the presence of magnetic moments, we examine how these
doublets are split by $\hat{H}_{ex}(\magDir)$ where $\magDir$ is
treated as an external parameter (related to classical magnetization;
single-Mn doped QDs where the Mn
magnetic moment behaves quantum-mechanically\cite{singleMn} require
different treatment) and represent them by an effective two-level 
Hamiltonian of Eq.~(\ref{eq:05}). We consider two example systems:
a simple four-level one where completely analytical treatment is
possible, and a more realistic envelope-function based model of a
cuboid QD.

\subsection{Four level model}

\begin{figure}
\includegraphics[width=7.5cm]{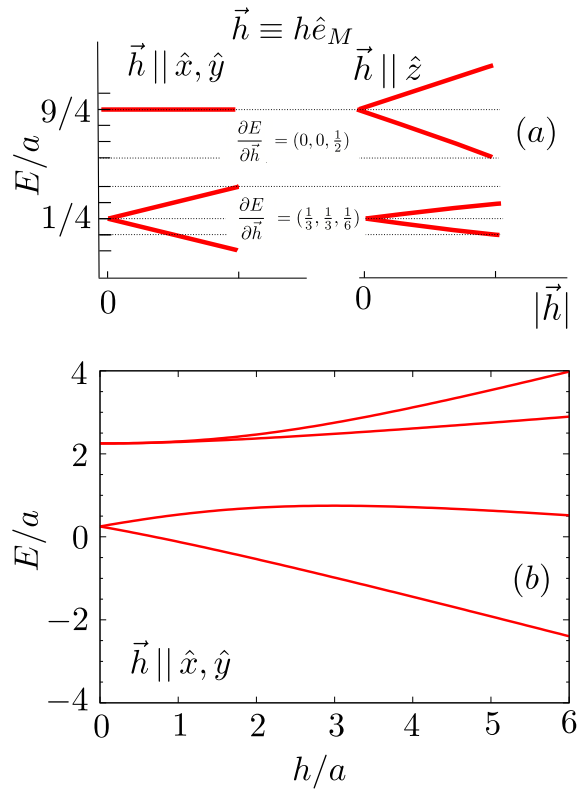}
\caption{(Color online) 
Splitting of levels $E(h)$ in a flat QD described by
Eq.~(\ref{eq:03}). (a) For the particular Kramers doublet, $E(h)$
depends on $\magDir$ and the $g$-factors (by convention
non-negative) are $\partial E/\partial\vec{h}=(g_x,g_y,g_z)$.  (b)
Beyond the linear regime in $h/a$, $\partial E/\partial (hn_x)$ will
be different for the upper and lower level of the split doublet, it
will depend on $h$ and may even change sign, indicating that the
$\Heff$ of Eq.~(\ref{eq:05}) based on parameters $g_{x,y,z}$
fails.}
\label{fig:01}
\end{figure}

Related to the Kohn-Luttinger Hamiltonian of a 
QW,\cite{Merkulov2010:bookCh,Kyrychenko2004:PRB}
the arguably simplest non-trivial model describing
anisotropy of a flat QD is
\begin{equation}\label{eq:03}
  \hat{H}_1=a\hat{J}_z^2+ \frac13 h\magDir\cdot \hat{\mathbf{J}}
\end{equation}
representing hole levels in a zinc-blende structure whose confinement
anisotropy and exchange splitting are parametrized by $a$ and $h$,
respectively (the term $a \hat{J}_z^2$ implies that the strongest
confinement is along the $z$-direction and this term also encodes information
about the SO coupling).
$\hat{J}_{x,y,z}$ are $4\times4$ spin-$\frac32$ matrices.
In terms of Eq.~(\ref{eq:01}), we now choose $\hat{H}=\hat{H}_1$
and the first (second) term in Eq.~(\ref{eq:03}) plays the role of 
$\hat{H}_{QD}$ ($\hat{H}_{ex}$).  Anisotropic behavior of eigenvalues
of $\hat{H}_1$, to linear order in $h/a$, is
illustrated in Fig.~\ref{fig:01}(a). It can be extracted from the
exact eigenvalues,
\begin{eqnarray}\label{eq:04a}
E_{hh}^\pm(h)&=&\frac54 a \pm \frac16 h +\sqrt{a^2+\frac19 h^2\mp \frac13 ah}\\
\label{eq:04b}
E_{lh}^\pm(h)&=&\frac54 a \pm \frac16 h -\sqrt{a^2+\frac19 h^2\mp \frac13 ah}
\end{eqnarray}
in the case $n_x=1$ (or $n_y=1$), 
shown in Fig.~\ref{fig:01}(b), which clearly differ 
from the case $n_z=1$ where the eigenvalues are strictly
linear functions of $h$ ($E_{hh}^\pm = 9a/4 \pm h/2$ and
$E_{lh}^\pm=a/4 \pm h/6$); subscripts refer to the
$E_{hh}^\pm(0)=9a/4$ (`heavy-hole', HH) and  $E_{lh}^\pm(0)=a/4$ 
(`light-hole', LH) doublets, respectively. In the limit of weak 
exchange, $h/a\ll 1$, splitting of each
of the Kramers doublets is symmetric and it can be characterized by
three parameters $|\partial E/\partial (hn_p)|$, $p=x,y,z$, for $h\to
0$ as depicted in Fig.~\ref{fig:01}(a).  These parameters can be
plausibly called, by analogy with the Zeeman effect, the anisotropic
$g$-factors $g_p$. From Eqs.~(\ref{eq:04a}),(\ref{eq:04b}), we
straightforwardly obtain 
$(g_x,g_y,g_z)=(0,0,1/2)$ and $(1/3,1/3,1/6)$ for the  
HH and LH doublet of the Hamiltonian $\hat{H}_1$, respectively. 
This result is known from the
context of QWs.\cite{Merkulov2010:bookCh,Petukhov1996:PRB}
We emphasize that these $g$-factors of the model specified by 
Eq.~(\ref{eq:03}) are
independent of the parameters $a,h$ (except for the requirement
$h\ll a$ which represents the $h\to 0$ limit).

If we focus on one particular Kramers doublet, it is straigtforward to
show that $\hat{H}_1$ projects to
\begin{equation}\label{eq:05}
  \Heff = 
  h\left[
 n_xg_{x}\hat\tau_x 
 +
 n_y g_{y}\hat\tau_y
 +
 n_zg_{z}\hat\tau_z 
  \right]
\end{equation}
for a suitably chosen basis $|K_1\rangle$, $|K_2\rangle$ of the
doublet. Here $\hat\tau_i$ are Pauli matrices and
we have mapped two eigenstates
of the original Hamiltonian $\hat{H}_{QD}$ on a pseudospin
$|\vec{\tau}|=1/2$ doublet $|+\rangle$, $|-\rangle$, where
$\hat{\tau}_z|\pm\rangle = \pm |\pm\rangle$. For
$\hat{H}=\hat{H}_1$, the eigenstates are only
four-dimensional (spanned by the $|J_z=3/2\rangle$,
$|J_z=-1/2\rangle$,$|J_z=1/2\rangle$,$|J_z=-3/2\rangle$
basis). We present another example of $\hat{H}$ in Sec.~IIB 
where advantage of the projection
becomes more apparent. The choice of basis $|+\rangle$, $|-\rangle$ is
crucial to obtain $\Heff$ in the simple form~(\ref{eq:05});
considering the HH
doublet: 
$|+\rangle=|J_z=3/2\rangle$, $|-\rangle=|J_z=-3/2\rangle$
leads to Eq.~(\ref{eq:05}) while for other basis choices the mapping
%
$\hat{H}_{ex}=
(h/3)\magDir\cdot\hat{\mathbf{J}}\ \mapsto \
  \Heff=h\magDir\cdot g\cdot \hat{\mathbf{\tau}}$
%
may lead\cite{Merkulov2010:bookCh} to non-symmetric tensor $g=g_{ij}$,
$i,j\in\{x,y,z\}$.  In general, if the mapping is to produce
$g_{ij}=\diag (g_x,g_y,g_z)$ the `suitable choice of the basis
$|K_1\rangle$, $|K_2\rangle$' where $|K_1\rangle\mapsto |+\rangle$ is
such that $\langle K_1|\hat{J}_{x,y}|K_1\rangle=0$, $\langle
K_1|\hat{J}_z|K_1\rangle\ge 0$ (and $|K_2\rangle$ is the time-reversed
image of $|K_1\rangle$ which is mapped to $|-\rangle$).



Let us now consider a general system described by Eq.~(\ref{eq:01}).
Assuming that the downfolding of $\hat{H}$ into $\Heff$ is 
possible for given $|K_1\rangle$, $|K_2\rangle$ (this assumption is
discussed in Appendix~A), the anisotropic $g$-factors
can readily be determined as $\partial E/\partial h$ for the
particular Kramers doublet level $E$. This is equivalent to 
perturbatively evaluating the effect of $\hat{H}_{ex}$ on two
degenerate levels to the first order of $h$ as follows: (i) specify
the Kramers doublet of interest, and find any basis 
$|K_1\rangle$, $|K_2\rangle$ of this doublet, (ii) extract the
operators $\hat{t}_{x,y,z}$ from $\hat{H}_{ex}$ by taking 
$\hat{t}_p=\partial \hat{H}_{ex}/\partial (n_p h)$ 
(for example, $\hat{t}_x=
\hat{J_x}/3$ for $\Heff$ appearing in $\hat{H}_1$), (iii) evaluate 
their matrices
%
\begin{equation}\label{eq:06}
  \tilde{t}_{x,y,z} = \left(\begin{array}{cc}
\langle K_1|\hat{t}_{x,y,z}|K_1\rangle & \langle K_1|\hat{t}_{x,y,z}|K_2\rangle 
\\
\langle K_2|\hat{t}_{x,y,z}|K_1\rangle & \langle K_2|\hat{t}_{x,y,z}|K_2\rangle 
  \end{array}\right)
\end{equation}
in the
two-dimensional space spanned by $|K_1\rangle$, $|K_2\rangle$,
and (iv) the non-negative eigenvalue of $\tilde{t}_p$ 
equals $g_p$ ($p=x,y,z$). 
We emphasize that while $g_p$ 
depends on system parameters in $\hat{H}_{QD}$ and
$\hat{H}_{ex}$, it also depends on which Kramers doublet we
choose. Higher doublets become relevant for QDs containing higher (odd) number
of holes, for example. 

The effective Hamiltonian in Eq.~(\ref{eq:05}) can be used for various
purposes, e.g., for studies of fluctuations of magnetization in magnetic
QDs\cite{Dorozhkin2003:PRB}, spin-selective tunneling through
non-magnetic QDs\cite{Katsaros2011:PRL} or excitons in single-Mn doped
QDs.\cite{LeGall2011:PRL}  If the magnetic easy axis is of interest,
the $g$-factors immediately provide the answer: $F(\magDir)$ based
on Eq.~(\ref{eq:05}) is minimized for $\magDir$ in the direction of the
largest $g_p$ (e.g. for the HH doublet in Fig.~\ref{fig:01}(a), it is
$n_z=1$ because $g_z>g_x,g_y$). If the full form of $F(\magDir)$ is
needed (e.g, for ferromagnetic resonance\cite{Liu2003:PRB}), 
it can be straightforwardly obtained by diagonalizing the
$2\times 2$ matrix of $\Heff$. Assuming $g_x=g_y$, 
the (modulus of the) eigenvalue can be expanded in terms of parameters
$A$ and $k$ as derived in Appendix~B.
It is meaningful to call 
\begin{equation}\label{eq:07}
A=(g_z^2-g_x^2)/(g_z^2+g_x^2)
\end{equation}
the asymmetry parameter since it vanishes in a perfectly cubic QD
($g_x=g_y=g_z$) and it is with respect to this parameter that we can
identify
\begin{eqnarray}\label{eq:08a}
  AF_1(\magDir) &=& -A k n_z^2 \\ \label{eq:08b}
  A^2F_2(\magDir) &=& +\frac18 A^2k(2n_z^2-1)^2
\end{eqnarray}
in Eq.~(\ref{eq:02}) to linear order of $k\propto h$.

\subsection{A cuboid quantum dot model}

With this general scheme at hand, we take one step in the hierarchy of
models towards a more realistic description of magnetic QDs. We
consider a zinc-blende structure p-doped semiconductor shaped into a
cuboid of size $L_x\times L_y\times L_z$ such as can be described by 
four-band Kohn-Luttinger Hamiltonian.\cite{Kyrychenko2004:PRB}
Also in this system, $\hat{H}=\hat{H}_2$ is a sum of
$\hat{H}_{ex}$ and $\hat{H}_{QD}$ but this time, 
$\hat{H}_{QD}$ comprises of blocks 
$\langle m_xm_ym_z|\hat{H}_{KL}| m_x'm_y'm_z'\rangle$ with
\begin{eqnarray}\nonumber
\hat{H}_{KL}&=&\frac{\hbar^2}{2m_0}\{\textstyle
    (\gamma_1+\frac52 \gamma_2) p^2  -
2\gamma_2[\hat{J}_x^2\hat{p}_x^2+\hat{J}_y^2\hat{p}_y^2+\hat{J}_z^2\hat{p}_z^2]
\\ && \label{eq:09}
-2\gamma_3[(\hat{J}_x\hat{J}_y+\hat{J}_y\hat{J}_x)\hat{p}_x\hat{p}_y
                                   +\mbox{c.p.}]
\end{eqnarray}
Here, $|m_xm_ym_z\rangle$ denotes the basis of envelope functions,
$\gamma_{1,2,3}$ the Luttinger parameters, $m_0$ the electron vacuum 
mass, $\hat{p}_{x,y,z}$ the momentum operators and c.p. denotes the cyclic
permutation (see Appendix~C for details). 
The envelope function is conveniently developed into
harmonic functions with $m_p-1$ nodes in the $p=x,y,z$ direction:
\begin{equation}\label{eq:10}
\langle \vec{r}|m_xm_ym_z\rangle =
N\sin \frac{m_x\pi x}{\lambda_xL}
       \sin \frac{m_y\pi y}{\lambda_yL}\sin \frac{m_z\pi z}{L}.
\end{equation}
We have introduced the dimensionless aspect ratios $\lambda_{x,y}=L_{x,y}/L$
and the normalization factor $N$. Our system can be viewed as an infinitely deep
potential well with $V(x,y,z)=0$ for $0<x<L_x$, $0<y<L_y$ and $0<z<L_z\equiv
L$ and infinite otherwise. 

\begin{figure}
\begin{tabular}{p{4.5cm}c}
\includegraphics[width=4.0cm,height=3cm]{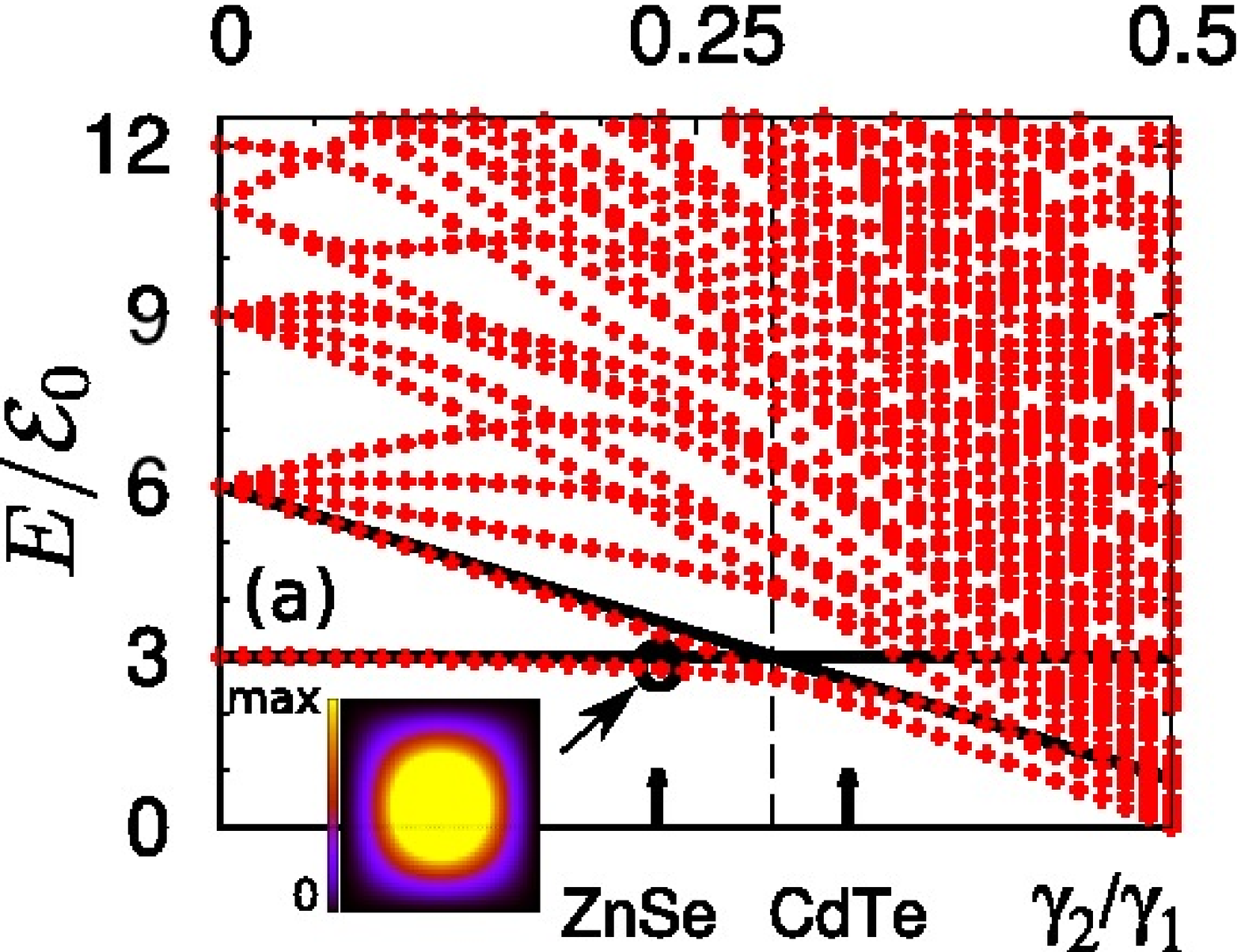}
\put(-48,12){\rotatebox{90}{\small $c_{R}$}}
\hbox{{\color{white}.}\hspace{-38mm}%
\includegraphics[scale=0.2,angle=-90]{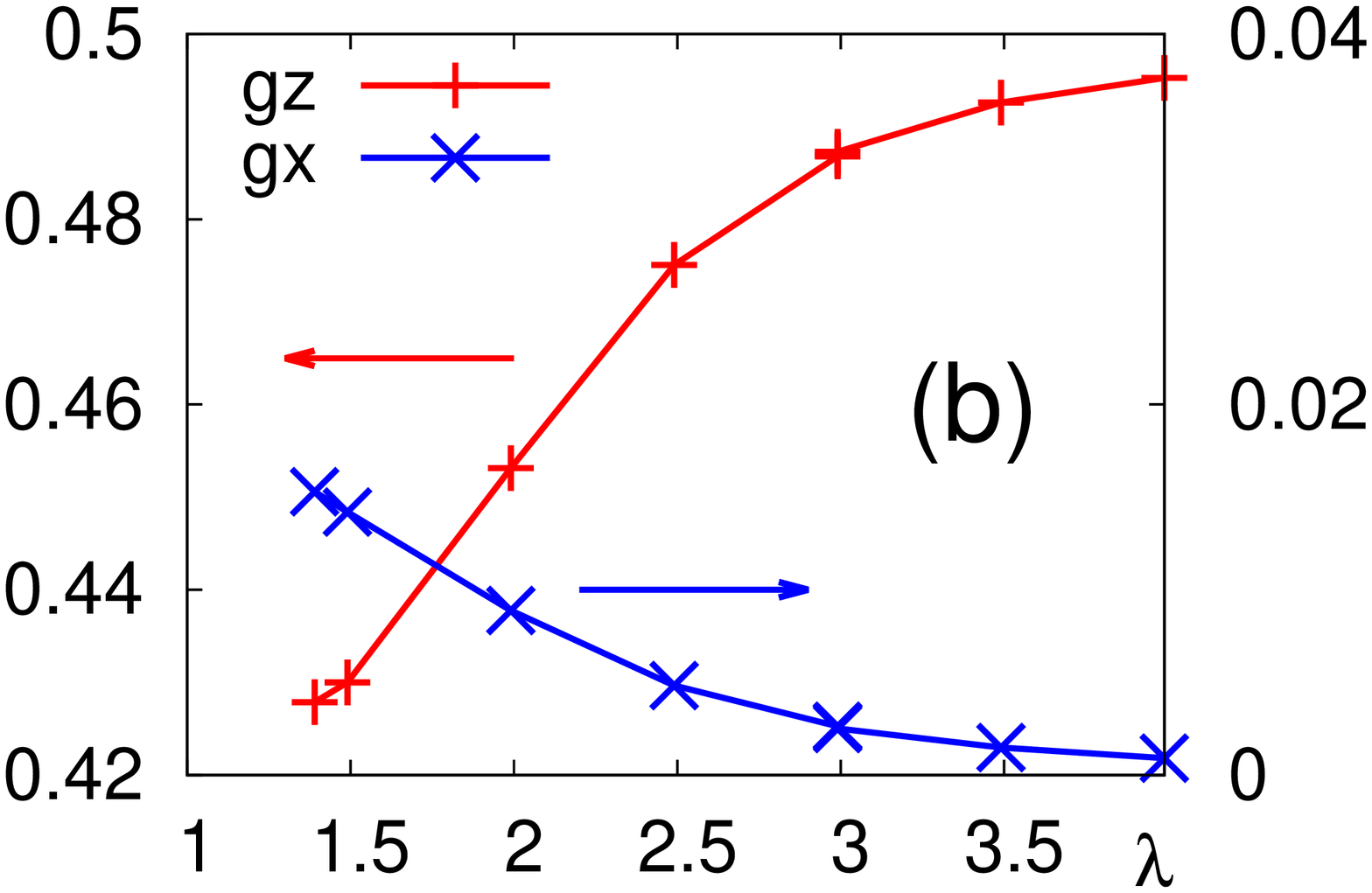}}
&
  $\begin{array}{ccc}  
  E \mbox{ [meV]} & g_x=g_y & g_z \\
71.7& 0.012  &     0.464  \\   
71.9& 0.305  &     0.171  \\
92.1& 0.167  &     0.160  \\
126.1& 0.274 &     0.237  \\ 
129.6& 0.076 &     0.069  \\
130.0& 0.082 &     0.045  \\
141.0& 0.205 &     0.212  \\
\\
49.9&    0.166&     0.164\\  
52.8&    0.027&     0.418\\
53.0&    0.269&     0.176\\
78.5&    0.162&     0.169\\
84.1&    0.010&     0.129\\
84.4&    0.064&     0.004\\
85.6&    0.203&     0.279
  \end{array}$
\end{tabular}
\unitlength=1mm
\put(-38,10){\rotatebox{90}{ZnSe}}
\put(-38,-20){\rotatebox{90}{CdTe}}
\caption{(Color online) 
  (a) Levels in a cubic dot (with $\gamma_3=\gamma_2$) in
  units of ${\cal E}_0$ defined by Eq.~(\ref{eq:11}). Solid lines
  indicate analytic result obtained when mixing between remote levels
  is disregarded. Note that their crossing
  (which we use to discern the weak and strong HH/LH splitting
  materials, dashed line) is very close to the actual crossing when
  level mixing is taken into account. Values representing ZnSe
  ($\bar{\gamma}_2/\gamma_1\approx 0.23$)  and CdTe
  ($\bar{\gamma}_2/\gamma_1\approx 0.33$) QDs are indicated. 
  Inset: squared wavefunction modulus of the ZnSe QD ground state in
  the $z=L/2$ section. (b) Dependence
  of the $g$-factors associated with the ground state Kramers doublet 
  in a CdTe QD on its shape ($\lambda_x=\lambda_y\equiv \lambda$).
  {\em Right:} Energies and $g$-factors in slightly deformed QDs
  ($\lambda=1.01$) for the lowest 7 Kramers doublets for ZnSe and CdTe,
  where ${\cal E}_0\approx 28\unit{meV}$ and $24\unit{meV}$,
  respectively, for $L=8\unit{nm}$.}
\label{fig:02}
\end{figure}

For fixed material parameters (Luttinger parameters in ratios
$\gamma_2/\gamma_1$, $\gamma_3/\gamma_2$) and QD shape
($\lambda_x,\lambda_y$), all matrix elements of 
all blocks $\langle m_xm_ym_z|\hat{H}_{KL}| m_x'm_y'm_z'\rangle$
scale as $1/L^2$. The spectrum, consisting of Kramers doublets which
occasionally combine into larger multiplets, is specified by a
sequence of dimensionless numbers $E/{\cal E}_0$ where
\begin{equation}\label{eq:11}
{\cal  E}_0=\hbar^2\pi^2\gamma_1/(2m_0L^2).
\end{equation}
For a cubic QD [$\lambda_x=\lambda_y=1$; see Fig.~\ref{fig:02}(a)] the
s-like state shown in the inset of Fig.~\ref{fig:02}(a) forms a
quadruplet, and depending on the value of $\gamma_2/\gamma_1$ (and to
somehow lesser extent also of $\gamma_3/\gamma_2$) this state competes
with the next doublet for having the lowest energy. The
critical value (see Appendix~C)
\begin{equation}\label{eq:12}
c_R=(2+128/9\pi^2)^{-1}\approx 0.29
\end{equation}
can be taken to distinguish materials with small
($\gamma_2/\gamma_1<c_R$, ground state quadruplet) and large
($\gamma_2/\gamma_1>c_R$, ground state doublet)
splitting between light and heavy holes in the bulk; these can be ZnSe and
CdTe, respectively, their values of $\bar{\gamma}_2/\gamma_1$ based on
approximating $\gamma_2$ and $\gamma_3$ by their average
$\bar{\gamma}_2=(\gamma_2+\gamma_3)/2$ are
indicated in Fig.~\ref{fig:02}(a). By numerical diagonalization we
have determined the lowest 7 Kramers doublets in slightly deformed QDs
($\lambda_x=\lambda_y\equiv \lambda = 1.01$) in these materials
($\gamma_{1/2/3}=4.8/0.67/1.53$ for ZnSe and $4.1/1.1/1.6$ 
for CdTe)\cite{matParam}
and executed the procedure (i)-(iv) above to obtain the
$g$-factors which are listed in the table on the right of
Fig.~\ref{fig:02} ($g_x=g_y$ due to $\lambda_x=\lambda_y$).
To avoid confusion, we remark that in (i), $|K_1\rangle$,
$|K_2\rangle$ are vectors of dimension 864 in the basis
$|m_xm_ym_z\rangle\otimes |J_z\rangle$ 
(see the discussion of cut-off in Appendix~C) 
and in (ii), $\hat{t}_x=(1/3) \hat{J}_x\otimes \eins_{xyz}$, where
$\eins_{xyz}$ is the identity operator in the space of the envelope
functions given by Eq.~(\ref{eq:10}). Evaluation and diagonalization
of the $2\times 2$ matrices in Eq.~(\ref{eq:06}) requested in (iii,iv)
is performed numerically. The possibility to map the action of
$\hat{H}_{ex}= (h/3)\magDir\cdot\hat{\mathbf{J}}\otimes\eins_{xyz}$ 
on the Kramers doublets $|K_1\rangle$, $|K_2\rangle$ implied by 
$\hat{H}_{QD}$ of a cuboid p-doped QD is discussed in Appendix~A.
 
The slight deformation of the QD makes the quadruplet split into two
doublets (with energies $71.7$ and $71.9\unit{meV}$ for ZnSe)
whose $g$-factors approach $(0,0,1/2)$ and $(1/3,1/3,1/6)$.
Similar situation occurs for the doublet pair with energies 
$52.8$ and $53.0\unit{meV}$ for CdTe. The actual ground
state in this material is, however, a doublet of different orbital character
than the quadruplet (we stress that this is due to the confinement, 
see Appendix~C); it evolves from the $E=6{\cal E}_0$ level of
$\gamma_2/\gamma_1=0$ as shown by the solid line in
Fig.~\ref{fig:02}(a) and its $g$-factors are isotropic,
$(1/6,1/6,1/6)$ in the limit $\lambda\to 1$. This doublet,
however, remains the ground state only in rather symmetric QDs 
($\lambda\approx 1.25$ in CdTe) and for more strongly deformed QDs,
the lower doublet of the $E=3{\cal E}_0$ (at $\gamma_2/\gamma_1=0$) quadruplet 
becomes the ground state just as it is the case for ZnSe for
arbitrarily small deformations $\lambda>1$. In Fig.~\ref{fig:02}(b),
we show how the $g$-factors of the CdTe QD ground state depend on
$\lambda$ beyond the mentioned value $\approx 1.25$.
These results, including the $g$-factors, are independent of the QD size $L$,
except for the energies which scale as $1/L^2$ as mentioned above.

From Fig.~\ref{fig:02}, one may conclude that the Ising-like
Hamiltonian is often an excellent approximation ($g_x=g_y=0$, as
others assume\cite{LeGall2010:PRB,Lee2000:PRB,Dorozhkin2003:PRB,%
Katsaros2011:PRL,LeGall2011:PRL}) for the lowest Kramers doublet.
To be more specific, we now
discuss materials with small and large HH/LH splitting separately. For
$\gamma_2/\gamma_1<c_R$, the out-of-plane $g$-factor ($g_z$)
overwhelmingly exceeds the in-plane one ($g_x=g_y$) even for minute
deformation of the QD; this can be seen from the numeric ZnSe data in
Fig.~\ref{fig:02}. We find $g_z=0.464$ and $g_x=g_y=0.012$ for
$\lambda-1$ as small as $0.01$. For CdTe, which represents the other
class ($\gamma_2/\gamma_1>c_R$), we find similar values ($g_z=0.418$)
for the {\em second} Kramers doublet while the lowest doublet remains
rather isotropic ($g_x=g_y=0.166$ and $g_z=0.164$). As we make the QD
deformation larger, these two doublets cross, so that the ground state
doublet is Ising like while the second lowest doublet 
remains more isotropic. This crossing occurs for $\lambda\approx 1.25$
in CdTe and data in Fig.~\ref{fig:02}(b) are only shown for
$\lambda>1.25$. 

We now elaborate on the properties of the low-energy sector of
$\hat{H}_2$ (at $h=0$). Coupling between blocks of
different $|m_x m_y m_z\rangle$ vanishes when 
$\gamma_3/\gamma_1,\gamma_2/\gamma_1\to 0$,
and Eq.~(\ref{eq:03}) becomes in this limit the exact effective
Hamiltonian of the lowest four levels ($m_p=1$ for all
$p=x,y,z$). They form a quadruplet for $\lambda=1$, which splits into
two doublets upon deformation of the QD; we can see it by writing
\begin{equation}\label{eq:13}
  \langle 111|\hat{H}_{KL}|111\rangle =  
  3{\cal E}_0 \left[\eins_4 f(\lambda) - 
              \hat{J}_z^2 (1-\lambda^{-2})\frac23 
              \frac{\gamma_2}{\gamma_1} \right]
\end{equation}
where $\eins_4$ is a unit $4\times 4$ matrix and $f(\lambda)$ 
is a certain function with $\lim_{\lambda\to 1}f(\lambda)=1$.
The lower doublet of this $4\times 4$ effective Hamiltonian 
has $g_z=1/2$ (when $\lambda>1$ and $\gamma_2>0$) 
and therefore the values of $g_z$ deviating from $0.5$
(appearing in Fig.~\ref{fig:02}) occur only due to admixtures from
higher-orbital ($m_p>1$) states of LH character. Indeed, going from
ZnSe to CdTe, the mixing becomes stronger and 
$g_z$ of the HH-like level drops from $0.464$ to $0.418$
($\lambda=1.01$, numerical data in Fig.~\ref{fig:02}). While 
Eq.~(\ref{eq:03}) may remain the effective Hamiltonian of the two
doublets originating from $|m_xm_ym_z\rangle=|111\rangle$ even for
$\gamma_2/\gamma_1>c_R$ (CdTe levels of $52.8$ and $53.0\unit{meV}$ in
Fig.~\ref{fig:02}), for $\lambda$ close to 1, there is the
more isotropic doublet on the stage ($49.9\unit{meV}$ in
Fig.~\ref{fig:02}). Nevertheless, if $\lambda$ is sufficiently large, 
the $\hat{J}_z^2$ term in Eq.~(\ref{eq:09}) will eventually dominate,
it will suppress all mixing between HH and LH states and the lowest
doublet will again 
approach $(g_x,g_y,g_z)=(0,0,0.5)$ as it is shown in Fig.~\ref{fig:02}(b).

\section{Magnetocrystalline anisotropy energy}

\begin{figure*}
\includegraphics[scale=0.5]{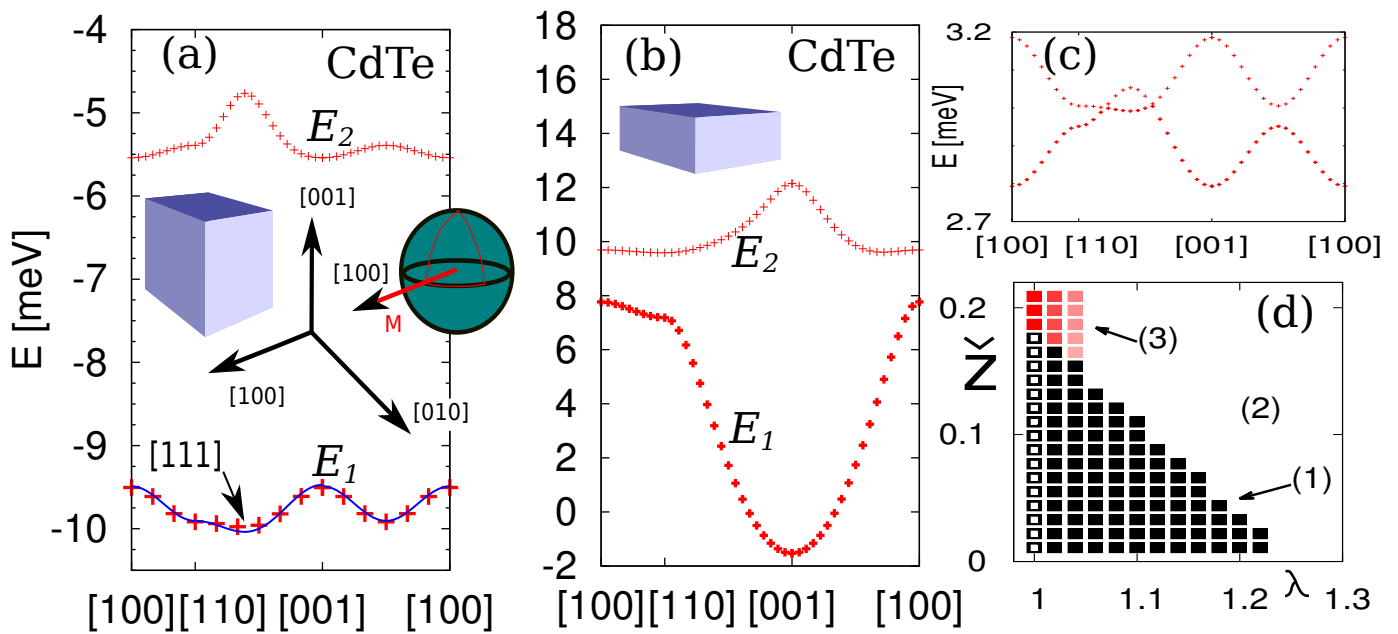}
\caption{(Color online) 
Magnetocrystalline energy as a function of magnetization
direction ($E_1$); the data labelled $E_2$ are explained in the text.
CdTe QD with $2.3\%$ Mn (a) $16\times 16\times 16\unit{nm^3}$, (b)
$16\times 16\times 8\unit{nm^3}$. (c) Fictitious material with parameters
described in the text; note that the sign of $K_c$ implied by
Eq.~(\ref{eq:14}) has changed compared to (a,b). 
(d) Color-coded easy axis positions for CdTe QDs as a function of
  aspect ratio ($\lambda$) and effective exchange splitting $\zz$. Black
  squares (1) indicate easy plane perpendicular to $z$-direction, 
  hollow squares denote an isotropic magnet; white region
  (2) corresponds to easy-axis [001]; red squares (3) denote systems
  with [111] easy axis which gradually shifts towards [001] with
  increasing $\lambda$. This plot is universal as far as $L$ is concerned.
}
\label{fig:03}
\end{figure*}

In analogy to the bulk systems, even cubic QDs retain
anisotropies. However, these cannot be described within the previous
framework: for instance, $g_x,g_y,g_z$ are all equal to $1/6$ in the
cubic CdTe QD ground state hence $A=0$ in Eq.~(\ref{eq:07}).
One could replace $g_{ij}$ by a higher rank tensor to
capture these effects, but MAE formalism of bulk magnets 
seems more customary and informative. 
Unlike the $g$-factors, MAE analysis does not invoke the concept of Kramers
doublets. The zero-temperature free energy $F(\magDir)$ 
of a magnetic QD with a single hole is now simply the lowest eigenvalue
of Eq.~(\ref{eq:01}) and it can be expanded in powers of 
$n_j$. The lowest terms compatible with cubic symmetry are\cite{note3}
\begin{equation}\label{eq:14}
  F_0 = K_c(n_x^4+n_y^4+n_z^4)+ 27K_{c2}n_x^2n_y^2n_z^2.
\end{equation}
For data calculated by numerically diagonalizing $\hat{H}=\hat{H}_2$ 
(model described in Sec.~IIB)
it turns out that Eq.~(\ref{eq:14}) suffices to obtain good fits;
for instance, lower solid line in Fig.~\ref{fig:03}(a) corresponds to 
$K_c=0.83\unit{meV}$ and $K_{c2}=0.075\unit{meV}$
with easy axis along $[111]$.
There we have chosen Cd$_{1-x}$Mn$_x$Te as the 
material, $L=16\unit{nm}$ and $h=50\unit{meV}$ which corresponds to 
$h=J_{pd}N_{Mn}S_{Mn}$ with $x\approx 2.3\%$ (we take\cite{matParam}
$|J_{pd}|=60\unit{meV\cdot nm^3}$, $S_{Mn}=5/2$ and
$N_{Mn}=4x/a_{l}^2$ with CdTe lattice constant $a_{l}=0.648\unit{nm}$).
Results in Fig.~\ref{fig:03} are again subject to scaling, similar to
the non-magnetic spectra in Fig.~\ref{fig:02}(a). When the material
parameters (specifically, $\gamma_2/\gamma_1$ and $\gamma_3/\gamma_2$)
are fixed, the spectrum of $\hat{H}_{2}$, expressed in the
units of ${\cal E}_0$, depends on a single dimensionless parameter
\begin{equation}\label{eq:15}
  \zz = h/{\cal E}_0 \equiv 2 m_0 h L^2/(\gamma_1 \pi^2\hbar^2).
\end{equation}
This scaling relates the spectra of e.g. cubic dots of different sizes and
Mn contents (if their respective values of $\zz$ are equal). Data in
Fig.~\ref{fig:03} therefore apply both to $x=2.3\%$ at $L=16\unit{nm}$
(if left as they are) and $x=9.2\%$ at $L=8\unit{nm}$ (if scaled by a
factor of 4). It turns out that the $g$-factor analysis presented in
the previous section is meaningful for $\zz\lesssim 0.1$ while now we
have stepped out of this limit. When the
exchange field $h$ becomes stronger, levels cross and
cease to depend linearly on $h$ as required by Eq.~(\ref{eq:05}); for
$\hat{H}=\hat{H}_1$, this is illustrated in Fig.~\ref{fig:01}(b). This
limit was determined for CdTe cubic QDs but it will typically not be
too different for other materials and/or aspect ratios $\lambda$ 
unless accidental (quasi)degeneracies occur at $\zz = 0$.

\begin{table}
\begin{tabular}{c|c|cc|ccccc}
       &       & \multicolumn{2}{c|}{cubic} && \multicolumn{4}{c}{deformed} \\
       &       & ZnSe  & CdTe  && \multicolumn{2}{c}{ZnSe} 
                                                   & \multicolumn{2}{c}{CdTe}\\
$h$ [meV]& $\zz$ &$K_c$ & $K_c$ && $K_c$   & $K_u$  &$K_c$    & $K_u$ \\  \hline
10     & 0.55  & 0.11  & 0.24  && 0.23    & -3.79  & 0.35     & -3.90 \\
20     & 1.1   & 0.20  & 0.43  && 0.36    & -5.52  & 0.62     & -6.21 \\
30     & 1.7   & 0.28  & 0.59  && 0.44    & -6.29  & 0.83     & -7.59 \\
40     & 2.2   & 0.35  & 0.71  && 0.50    & -6.72  & 1.01     & -8.56 \\
50     & 2.8   & 0.41  & 0.83  && 0.56    & -7.01  & 1.16     & -9.30 

\end{tabular}
\caption{Magnetic anisotropy constants (in meV) for a $16\times 16\times 16
\unit{nm^3}$ (cubic) and $16\times 16\times 8\unit{nm^3}$ (deformed)
ZnSe and CdTe magnetic QD as a function of exchange splitting (or dimensionless
parameter $\zz$ as for CdTe). }\label{tab:01}
\end{table}

MAE shown in Fig.~\ref{fig:03} describe systems well beyond this limit of small
$\zz$ (linear regime). We first focus on a perfectly cubic CdTe QD where
there are no anisotropies in the linear regime. As already mentioned, the
lowest energy hole state in Fig.~\ref{fig:03}(a) exhibits a [111] easy
axis with $K_c=0.83\unit{meV}$ at 
$L=16\unit{nm}$ and $h=50\unit{meV}$, i.e. $\zz\approx 2.8$, 
(this corresponds to a realistic $x\approx 2.3\%$ Mn
doping). In bulk DMSs, [111] would be an unusual magnetic easy axis
direction\cite{Zemen2009:PRB} and we surmise that the reason for this
is that for instance in  (Ga,Mn)As grown on a GaAs substrate,
there is a sizable compressive strain which prefers either parallel or
perpendicular orientation of $\magDir$ with respect to the growth axis.

We note that in a QD containing two holes (closed-shell
system\cite{Oszwaldowski2011:PRL}) the anisotropies will also be
present and they will be different from the single-hole case.  Free
energy, taken as a sum, $F_0(\magDir)=E_1+E_2$, of the lowest two
single-hole states [shown e.g. in Fig.~\ref{fig:03}(a)], is not a
constant independent of $\magDir$ as one could naively expect. This
intuition reflects $\Heff$ in Eq.~(\ref{eq:05}) where the two hole
states have opposite spin (hence their energies add up to zero). Once
we leave the linear regime ($\zz \gtrsim 0.1$), $\Heff$ ceases to be a
good approximation.  Qualitatively, the same behaviour is found for
ZnSe (not shown), a smaller value of $K_c=0.41\unit{meV}$ is accounted
for by the smaller HH/LH splitting. The value of this constant is a
complicated function of system parameters and it can even change sign
as shown in Fig.~\ref{fig:03}(c) where $K_c=-0.63\unit{meV}$.
Parameters used in this figure
($\gamma_1/\gamma_2/\gamma_3=4.0/1.5/1.6$ and $h=20\unit{meV}$) do not
strictly correspond to published values of any semiconductor but they
can be viewed as reasonable given the uncertainty in experimental
determination of the Luttinger parameters.  Dependence of the
anisotropy constants for ZnSe and CdTe QDs on $h$ is summarized in
Tab.~\ref{tab:01}.

Let us now return to non-cubic QDs. As already explained, the sizable
$g$-factor anisotropies shown in Fig.~\ref{fig:02}(b), relevant to the
case of weak magnetism ($\zz\ll 1$), translate into an additional term
$AF_1=K_u n_z^2$ in the free energy of Eq.~(\ref{eq:02}) where $K_u=-kA$ up
to linear order in $\zz\propto k$. Typically, $K_u$ exceeds $K_c$
already for small QD deformation ($\lambda$ slightly over one) and the
data in Fig.~\ref{fig:03}(b) imply $K_u$ almost an order of magnitude
larger than $K_c$ for $\lambda=2$ (see also data in Fig.~\ref{fig:02}
where $g_z\gg g_x$). Regardless of the contributions to $K_u$ of
higher order in $\zz$, data in Tab.~\ref{tab:01} imply an out-of-plane
easy axis (in the [001] direction) as it is the case in QWs. 
However, upon deforming of a QD the easy axis does not abruptly
jump from [111] to [001] but smoothly interpolates between these two
directions. Similar effect, easy axis shifting as a function of some
system parameter, is also known in bulk DMS [(Ga,Mn)As epilayers in
  particular, see Fig.~8 in Ref.~\onlinecite{Zemen2009:PRB}]. Easy
axes as a function of QD shape (oblate dots, $\lambda>1$) and
effective exchange splitting $\zz$ are summarized in
Fig.~\ref{fig:03}(d) and the mentioned gradual shift of easy axis is
indicated by shading between regions (3) and (2) (easy axes $[111]$
and $[001]$, respectively).  On the other hand, the easy axis position
changes abruptly between (1) and (3) or (1) and (2); region (1)
corresponds to easy axis in the plane perpendicular to $[001]$ (with
anisotropies within this plane being very small). The abrupt changes
reflect ground state crossings, such as the one with $\lambda$
described below Eq.~(\ref{eq:12}), while the gradual ones stem from
level mixing caused by $\hat{H}_{ex}$.

Finally, we comment on MA in QDs occupied by more than one hole. As
already mentioned above, one possible approach is to discuss
open-shell and closed-shell systems separately. This notion is based
on the concept of the QD being an artificial atom whose
levels are organized into shells comprising of spin-up and spin-down
orbitals. Whenever a shell is completely filled (closed), the numbers
of spin-up and spin-down carriers are equal hence their total spin is
zero. If the QD is magnetically doped, no magnetic ordering is
expected and also no MA. However, strong SO coupling puts this
concept into question since it mixes different shells and also
invalidates the spin-up and down labels of individual orbitals. The MA
as a function of particle number $N_p$ strongly varies, both quantitatively
and qualitatively. By comparing the $N_p=1$ and $N_p=2$ cases of a cubic
CdTe QD, that is $F_0(\magDir)=E_1$ and $F_0(\magDir)=E_1+E_2$ of
Fig.~\ref{fig:03}(a), we find that while the easy axis $[111]$ 
in the former case
is relatively `soft' (energy difference between $\magDir||[111]$ and
$[110]$ is `only' $\approx 0.1\unit{meV}$), the QD with two holes has
a `robust' easy axis $[110]$ and the corresponding minimum in
$F_0(\magDir)$ is as deep as $0.3\unit{meV}$. MA as a function of $N_p$
displays rich behavior and one can therefore envision control of
nanomagnetism by electrostatic gating, illumination (used to photoinduce
carriers) and possibly also temperature, known to alter the magnetic
ordering in the bulk-like structures.\cite{Zutic2004:RMP,igor}

\section{Conclusions}

We have discussed two approaches to magnetic anisotropies in quantum
dots (QDs) described by a generic model in Eq.~(\ref{eq:01}). An effective
Hamiltonian for individual Kramers doublets allows to express the
energetics of a magnetically doped QD in terms of only three
parameters (anisotropic $g$-factor) if the exchange splitting due to
the magnetic ions is relatively small. On the other hand, if the exchange
splitting is large or the QD's symmetry is too high, the
symmetry-based expansion of the magnetocrystalline energy in powers of
the direction cosines of magnetization may in principle contain
infinitely many terms (each of them quantified by one
parameter). Focusing on manganese-doped semiconductor QDs, we find
that only first few terms are appreciable, present their values and
show in Fig.~\ref{fig:03}(d) a diagram of magnetic anisotropies in the
QD parameter space. While we focus on a relatively
small parameter range in that diagram, and the barriers
between individual free energy minima are relatively low, it
demonstrates that the QDs may have rich magnetic anisotropies. In
spintronics,\cite{Zutic2004:RMP,Fabian2007:APS,Tarucha2006:bookCh} 
these systems could thus enable confinement-controlled multi-level
logic.  Our results provide a starting point for further studies of
nanoscale magnetism in QDs. Such studies could relax the mean-field
approximation, include multiple-carrier
states,\cite{fewElectrons,Cheng2009:PRB} or the effect of strain.

\section*{Acknowledgements}

This work is supported by DOE-BES DE-SC0004890, NSF-DMR 0907150, 
AFOSR-DCT FA9550-09-1-0493, U.S. ONR N0000140610123, 
and NSF-ECCS 1102092. 

\section*{Appendix A}

The downfolding of $\hat{H}_{QD}+\hat{H}_{ex}$ to $\Heff$ is indeed
possible for the two example systems discussed in Sec.~IIA and~IIB.
To prove this, we first transform the basis $|K_1\rangle,|K_2\rangle$
to $|K_1'\rangle,|K_2'\rangle$ where $\tilde{t}_z$ of
Eq.~(\ref{eq:06}) is diagonal and then verify that the diagonal
elements of $\tilde{t}_x$ and $\tilde{t}_y$ vanish. This procedure has
to be applied to each Kramers doublet of interest. In the case of
$\hat{H}_1$ in Eq.~(\ref{eq:03}), this is done simply by construction
(e.g. $|K_1'\rangle,|K_2'\rangle$ for the upper doublet in
Fig.~\ref{fig:01}(a) is just $|J_z=3/2\rangle,
|J_z=-3/2\rangle$).  In the model described by $\hat{H}_2$,
one can split the Hilbert space into two disjunct subspaces
$\mathscr{H}_1$, $\mathscr{H}_2$ and the above assertion can be shown
to hold if $|K_1'\rangle\in\mathscr{H}_1$ and
$|K_2'\rangle\in\mathscr{H}_2$. (The decomposition
$\mathscr{H}_1\oplus\mathscr{H}_1$ relies on $\hat{H}_{ex}$ being
independent of space coordinates; relaxing the mean-field treatment of
Mn magnetic moments thus introduces corrections to
$\hat{H}_{\mathrm{eff}}$.) Finally, one adjusts the relative phase
between $|K_1'\rangle$ and $|K_2'\rangle$, so that the matrix
$\tilde{t}_x$ is real and $\tilde{t}_y$ purely imaginary.

\section*{Appendix B}

This Appendix explains the relation between the anisotropic
$g$-factors and Eq.~(\ref{eq:02}).
The eigenvalues of $\Heff$ are two numbers
of equal magnitude and opposite sign, the lower of which is
\begin{equation}\label{eq:16}
-h\sqrt{n_x^2g_x^2+n_y^2g_y^2+n_z^2g_z^2}.
\end{equation}
Let us consider for example single hole in a cuboid QD of dimensions
$\lambda L\times \lambda L\times L$ (such as it corresponds to data in
Fig.~\ref{fig:02}) so that $g_x=g_y$. Expression~(\ref{eq:16}) which
now equals $F(\magDir)$ can be rewritten as
\begin{equation}\label{eq:17}
  -\frac{h\sqrt{g_x^2+g_z^2}}{\sqrt{2}}
  \sqrt{1+\frac{g_z^2-g_x^2}{g_z^2+g_x^2} (n_z^2-n_x^2-n_y^2)}
\end{equation}
and developped in terms of a small parameter $A=(g_z^2-g_x^2)/(g_z^2+g_x^2)$
which quantifies the QD asymmetry as
\begin{equation}\label{eq:18}
  {\textstyle -k(1-\frac12 A)-A k n_z^2+\frac18 A^2k(2n_z^2-1)^2+\ldots}
\end{equation}
where $k=h\sqrt{(g_x^2+g_z^2)/2}$. The first term does not depend on
the magnetization direction, hence it can be disregarded for the
purposes of magnetic anisotropy analysis.

\section*{Appendix C}

We derive Eq.~(\ref{eq:12}) in this Appendix and discuss the details
of the model considered in Sec.~IIB. Energies $E/{\cal E}_0$
in Fig.~\ref{fig:02}(a) are calculated by numerical diagonalization of
$\hat{H}_2$ with $h=0$,
a matrix constructed of $4\times 4$ blocks $\langle m_xm_ym_z|\hat{H}_{KL}|
m_x'm_y'm_z'\rangle/{\cal E}_0$ introduced at the beginning of Sec.~IIB.
The basis of $\hat{H}_{QD}$ consists thus of direct product states
$|m_xm_ym_z\rangle\otimes |J_z\rangle$ where $|J_z\rangle$ are the
four-spinors of total angular momentum $J=3/2$ which are eigenstates
to $\hat{J}_z$. For practical purposes, we cut-off the basis by
$m_x,m_y,m_z\le 6$, resulting in $\hat{H}_{QD}$ of dimension
864. Eigenvalues are typically converged to better than
$0.1\unit{meV}$ for this cut-off.

The matrix $\hat{H}_{QD}/{\cal E}_0$ is block-diagonal for 
$\gamma_2=\gamma_3=0$
and the block $m_x,m_y,m_z$ has a four-fold degenerate eigenvalue
\begin{equation}\label{eq:19}
  (m_x/\lambda_x)^2+(m_y/\lambda_y)^2+m_z^2.
\end{equation}
Dimensionless energies on the left of Fig.~\ref{fig:02}(a) correspond to
$\lambda_x=\lambda_y=1$ and are hence integers. The lowest level
$E/{\cal E}_0=3$ belongs to $(m_x,m_y,m_z)=(1,1,1)$ while the first
excited state $E/{\cal E}_0=6$ entails an additional threefold
geometric degeneracy corresponding to orbital states $(1,1,2)$,
$(1,2,1)$ and $(2,1,1)$; the $E/{\cal E}_0=6$ level for
$\gamma_2=\gamma_3=0$ is thus twelve-fold degenerate. 

Next, we can treat the HH-LH splitting as a perturbation when
$\gamma_2$ and $\gamma_3$ are turned on. In the lowest order, 
mixing between different $(m_x,m_y,m_z)$ blocks can be neglected
except for the case when their energies were equal at
$\gamma_2=\gamma_3=0$ as in the case of the three blocks of the 
$E/{\cal E}_0=6$ level. With coupling to the remote levels
disregarded, we are left with a $12\times 12$ matrix in this case
which can be diagonalized analytically. It turns out to have two
four-fold degenerate eigenvalues
\begin{equation}\label{eq:20}
  E_4^\pm/{\cal E}_0=6+\frac{64}{3\pi^2} \frac{\gamma_2}{\gamma_1}
  \left(s\pm \sqrt{s^2+\frac{81\pi^4}{1024}}\right)
\end{equation}
and two two-fold degenerate ones
\begin{equation}\label{eq:21}
  E_2^\pm/{\cal E}_0=6-\frac{128}{3\pi^2} \frac{\gamma_2}{\gamma_1}
  \left(s\mp \frac{9\pi^2}{64}\right).
\end{equation}
The lowest of these four energies is $E_2^-$ and it is shown in
Fig.~\ref{fig:02}(a) for $s\equiv \gamma_3/\gamma_2=1$ as a solid line
which crosses the horizontal line $E/{\cal E}_0=3$ corresponding to
the $(m_x,m_y,m_z)=(1,1,1)$ quadruplet which does not shift in energy
to the first order of this perturbation analysis. Eq.~(\ref{eq:12})
is the solution of $E_2^-=3{\cal E}_0$ for $\gamma_2/\gamma_1$ under
the assumption $s=1$. Such level crossing (as a function of
$\gamma_2/\gamma_1$) is genuinely due to the confinement and no level
crossings occur in in bulk as long as $0<\gamma_2/\gamma_1<1/2$.

\end{document}